\documentclass[aps,pra,twocolumn,noshowpacs,noshowkeys,amsmath,amssymb,authoryear,nofootinbib]{revtex4}
\usepackage{graphicx}
\pdfoutput=1
\usepackage{bm}
\usepackage [autostyle, english = american]{csquotes}
\MakeOuterQuote{"}
\usepackage{dcolumn}
\usepackage{xcolor}
\usepackage[colorlinks,citecolor=blue,urlcolor=blue,bookmarks=false,hypertexnames=true,linkcolor=blue]{hyperref} 
\usepackage{amsmath}
\usepackage[skip=0.333\baselineskip]{caption}
\usepackage{subcaption}

\def\noi{\noindent}
\def\bc{\begin{center}}
\def\ec{\end{center}}
\newcommand{\bea}{\begin{equation}}
\newcommand{\eea}{\end{equation}\noi}
\newcommand{\ber}{\begin{eqnarray}}
\newcommand{\eer}{\end{eqnarray}\noi}
\begin{document}
\title{Applying the Nash Bargaining Solution for a Reasonable Royalty II}
\author{David M. Kryskowski$^{1}$}\email{davidkryskowski@gmail.com}
\thanks{The views expressed in this paper do not necessarily represent the views of the Arizona Department of Revenue. }
\author{David Kryskowski$^{2}$}
\affiliation{$^{1}${Arizona Department of Revenue, Phoenix, Arizona, 85007, USA}\\
$^{2}$UD Holdings, 2214 Yorktown Dr., Ann Arbor, Michigan, 48105, USA}

\date{July 19, 2024}

\begin{abstract}
This paper expands on the concepts presented in \emph{Applying the Nash Bargaining Solution for a Reasonable Royalty}. The goal is to refine the process for determining a reasonable royalty using statistical methods in cases where there is risk and uncertainty regarding each party's disagreement payoffs (opportunity costs) in the Nash Bargaining Solution (NBS). This paper uses a Bayes Cost approach to analyze Case 1, Case 2, and the Original Nash model from the authors' previous work. By addressing risk and uncertainty in the NBS, the NBS emerges as a more reliable method for estimating a reasonable royalty, aligning with the criteria outlined in \emph{Georgia Pacific} factor fifteen. 
\end{abstract}




\maketitle

\section{Introduction}

Determining a reasonable royalty for intellectual property infringement remains a pivotal issue in patent litigation. As defined by U.S. law\footnote{35 U.S.C. § 284.}, a reasonable royalty is often the basis for awarding damages in such cases. The complexity of accurately assessing what constitutes a "reasonable" royalty introduces significant challenges stemming from the inherent subjectivity and varied interpretations of fairness and equity within licensing agreements \citep{jarosz2012hypothetical}. The U.S. judicial system has articulated several factors to guide the determination of reasonable royalties, notably through the \textit{Georgia Pacific}\footnote{\emph{Georgia-Pacific Corp. v. U.S. Plywood Corp.}, 318 F. Supp. 1116, 1120 (S.D.N.Y. 1970), mod. and aff’d, 446 F.2d 295 (2d Cir. 1971), cert. denied, 404 U.S. 870 (1971).} factors. One of these factors, \textit{Georgia Pacific} factor fifteen\footnote{Factor fifteen is: “The amount that a licensor (such as the patentee) and a licensee (such as the infringer) would have agreed upon (at the time the infringement began) if both had been reasonably and voluntarily trying to reach an agreement; that is, the amount which a prudent licensee -- who desired, as a business proposition, to obtain a license to manufacture and sell a particular article embodying the patented invention -- would have been willing to pay as a royalty and yet be able to make a reasonable profit and which amount would have been acceptable by a prudent patentee who was willing to grant a license.”}, allows for a hypothetical negotiation to establish a reasonable royalty, suggesting that bargaining models like the Nash Bargaining Solution (NBS) can be used as a tool to aid in determining a royalty.

Developed by economist John Forbes Nash, the NBS aims to optimize the distribution of gains from bargaining in cooperative scenarios \citep{nash1950bargaining}. The NBS relies on the parties' disagreement payoffs (opportunity costs) and bargaining weight \citep{binmore1986nash}. Employing the NBS is not without challenges. The primary difficulty is determining the inputs the model requires, such as each party's disagreement payoffs. These inputs are frequently obscured by limited data availability and case-specific factors. Despite these challenges, the NBS provides a foundation for negotiating a royalty that can reflect the true economic value, considering the risks and uncertainties associated with these inputs.

This paper builds on the authors' previous work by aiming to refine the process of determining reasonable royalties under conditions of risk and uncertainty \citep{friedman1948utility,robert2007bayesian}. By integrating estimation theory and adopting cost functions to the specificities of intellectual property negotiations, this work offers a practical approach to capturing the complexities inherent in such negotiations. Through the analysis of hypothetical cases and a review of adaptations to the Original Nash model presented in earlier work, \emph{Applying the Nash Bargaining solution for a Reasonable Royalty} \citep{kryskowski2022applying}, this paper seeks to advance a reliable method for estimating reasonable royalties that align with judicial criteria and economic reality. 

The outcome of this paper is intended to contribute to the broader academic and professional discourse by providing a clearer, more quantifiable basis for applying the NBS in legal settings, thereby aligning economic theory with practical, actionable legal strategies.

\section{Nash Bargaining Solution and Reasonable Royalties}
\label{sec:models}

The NBS is a pivotal theoretical framework in negotiating intellectual property royalties. Originating from the seminal work of John Forbes Nash, it provides a systematic approach for determining the optimal distribution of gains from cooperative bargaining scenarios. This section explores the NBS's foundational concepts, its application in intellectual property rights, and the enhancements necessary to adapt it to contemporary legal and economic landscapes.

\subsection{Theoretical Foundations of the Nash Bargaining Solution}
\label{sec:foundations}

John Forbes Nash introduced the NBS as a solution concept in cooperative game theory \citep{nash1950bargaining,nash1953two}, characterized by several axioms that define an ideal negotiation outcome:

\begin{enumerate}
\item{\textbf{Individual rationality:}  No party will agree to accept a payoff lower than the one guaranteed to him under disagreement.}
\item{\textbf{Pareto optimality:} None of the parties can be made better off
without making at least one party worse off.}
\item{\textbf{Symmetry:} If the parties are indistinguishable, the agreement should not discriminate between them.}
\item{\textbf{Affine transformation invariance:} An affine transformation of the payoff and disagreement point should not alter the outcome of the bargaining process.}
\item{\textbf{Independence of irrelevant alternatives:} All threats the parties might make have been accounted for in the disagreement point.}

\end{enumerate}

In the realm of intellectual property, the practical application of the NBS is guided by the criteria outlined in the \textit{Georgia Pacific} case, particularly factor fifteen. This factor anticipates a hypothetical negotiation scenario to determine a reasonable royalty. The NBS aptly fits this scenario by considering the potential payoffs each party would receive without an agreement, thus forming a basis for reasonable royalty calculations.

\subsection{Background on \emph{Applying the Nash Bargaining Solution for a Reasonable Royalty}}

In the authors' previous paper \citep{kryskowski2022applying}, the Normalized Royalty Model was introduced, influenced by Choi and Weinstein's Two Supplier World Model \citep{choi2001analytical}. The Normalized Royalty Model normalizes all monetary terms between zero and one, simplifying the determination of royalties using financial statement variables and making the NBS more accessible for courts and juries in litigation.

Additionally, the previous paper defined the bargaining weight in the NBS by introducing the perception equation found in Section \ref{sec:Bargaining Weight and Perception of Strength}. Typically, royalty negotiations assume equal bargaining power or a 50/50 split of the surplus between parties\footnote{\textit{VirnetX, Inc. v. Cisco Systems, Inc.}, 767 F.3d 1308 (Fed. Cir. 2014); \textit{Oracle Am., Inc. v. Google Inc.}, 798 F. Supp. 2d 1111 N.D. Cal. 2011; \textit{Illumina, Inc. v. BGI Genomics Co., Ltd}, No. 19-CV-03770-WHO, 2021 WL 4979799 (N.D. Cal. Oct. 27, 2021).}. However, by defining the bargaining weight, the NBS can be used symmetrically or asymmetrically, allowing for estimating the parties' relative bargaining weights based on the facts of the case. The technique of using the perception equation allows for the use of the NBS without the drawback the Federal Circuit identified in \emph{VirnetX}\footnote{It is recommended that the reader knows the background of the NBS as it relates to this case.} because it changes the starting assumption of equal bargaining power\footnote {See "Profit Split" from {\textit{Contour Ip Holding, LLC,} No. 3:17-CV-04738-WHO, 2021 WL 75666 (N.D. Cal. Jan. 8, 2021).}}.

\subsection{Literature on the NBS and Reasonable Royalties}

The literature on the asymmetric NBS and its application in intellectual property litigation offers valuable insights. Bhattacharya's work \citep{bhattacharya2019nash} explores the "Nash Program," which aims to bridge cooperative and non-cooperative game-theoretic models \citep{serrano2021sixty}, illustrating asymmetric bargaining power in the \emph{VirnetX} case. Wright and Yun \cite{wright2019use} write about various instances in which bargaining models were used in litigation. Furthermore, Kankanhalli and Kwan's study \citep{kankanhalli2024bargaining} delves into the sources of bargaining power in royalty negotiations, enhancing understanding of how bargaining power impacts royalty allocations. They support the view that surplus should be allocated asymmetrically based on the relative bargaining power between the parties, as also suggested by Sidak \citep{sidak2015bargaining}\footnote{See \protect\citep{kankanhalli2024bargaining} footnote 10.}. Additionally, Zimmerck's work  \citep{zimmeck2011game} is notable for building NBS models where the parties do not have equal bargaining power, further contributing to this field of study. Notably, Reed-Arthurs, Akemann, and Teece give an excellent overview of the legal landscape for using bargaining models and a detailed review of the Rubinstein model \citep{reed2021resolving}.

\subsection{Normalized Royalty Model}

The inputs to Normalized Royalty Model include the operating income for the licensor ($\pi_1$), the operating income for the licensee ($\pi_2$), the operating revenue ($OR$), operating cost ($OC$), operating income ($OI$=$OR$-$OC$), operating margin ($OM$=$OI/OR$), the royalty rate ($r$), the disagreement payoffs ($d_1$ for licensor and $d_2$ for licensee), and the bargaining weight ($\alpha$). 

In this context, party 1 represents the licensor, while party 2 represents the licensee. For a comprehensive derivation of the NBS and detailed definitions of the financial variables, refer to \citep{kryskowski2022applying} and \citep{vernimmen2022corporate}.

The optimal partition of profits under the asymmetric NBS in this model can be expressed as \citep{kalai1977nonsymmetric,binmore1986nash,roth2012axiomatic}:

\begin{subequations}\label{eq:NBTS1}
\begin{align}
\pi_1^* &= d_1 + \alpha \left( O_I - d_1 -d_2 \right) \label{sub-eq-1:1} \\ 
\pi_2^* &= d_2+ (1-\alpha) \left( O_I - d_1 -d_2 \right) \label{sub-eq-2:1} 
\end{align}
\end{subequations}

Under the Normalized Royalty Model, the payoffs for parties 1 and 2, respectively, are:

\begin{equation}
\label{eq:NBTS2}
\frac{\pi_1^*}{O_I} =  \frac{r  \, O_R}{O_I} = \frac{r}{O_M}
\end{equation}

\noindent
\begin{equation}
\label{eq:NBTS3}
\frac{\pi_2^*}{O_I} =  \frac{O_R-O_C-rO_R}{O_I} = 1-\frac{r}{O_M} 
\end{equation}

\noindent
Defining the normalized variables:

\begin{equation}
\label{eq:NBTS4}
d_1^\dagger = \frac{d_1}{O_I}  ~~~~~ 0 \leq d_1^\dagger \leq 1
\end{equation}

\noindent

\begin{equation}
\label{eq:NBTS5}
d_2^\dagger = \frac{d_2} {O_I}   ~~~~~ 0 \leq d_2^\dagger \leq 1
\end{equation}

\noindent

Substituting Eqs. \eqref{eq:NBTS2} and \eqref{eq:NBTS4}--\eqref{eq:NBTS5} into Eq. \eqref{sub-eq-1:1}, the optimal royalty is obtained with an arbitrary bargaining weight for party 1 is:

\noindent

\begin{equation}
\label{eq:NBTS6}
\frac{r}{O_M}= d_1^\dagger + \alpha  \left(1 - d_1^\dagger - d_2^\dagger \right)
\end{equation}

Where:
\begin{equation}
\label{eq:NBTS7}
0 \leq d_1^\dagger + d_2^\dagger \leq 1
\end{equation}

\hspace{1cm}

Throughout this paper, the normalized royalty, $\frac{r}{O_M}$, is referred to as $\theta$.

\subsection{Bargaining Weight and Perception of Strength}
\label{sec:Bargaining Weight and Perception of Strength}

The bargaining weight is critical in determining how surplus is distributed between negotiating parties, reflecting their respective bargaining strengths. While an equal split of surplus at $\alpha = 1/2$ is a common assumption, using the perception equation to define $\alpha$ offers a more nuanced perspective.

The perception equation formalizes the bargaining weight, considering each party's own bargaining strength and how they perceive each other's bargaining strength. It introduces the parameter $P_{m,n}$, representing party m's perceived bargaining strength by party n. Assuming that the bargaining strength of each party is the average of their own perception and the perception of the other party, the following ansatz is used to describe the bargaining weight of party 1:

\begin{subequations}\label{eq:NBTS17}
\begin{align}
\alpha_1 &= \frac{1}{2}\left[P_{1,1} + P_{1,2}\right] \label{sub-eq-1:3} \\
\alpha_2 &= 1-\frac{1}{2}\left[P_{2,1} + P_{2,2}\right] \label{sub-eq-2:3}
\end{align}
\end{subequations}

By averaging Eqs. \eqref{sub-eq-1:3} and \eqref{sub-eq-2:3}, the complete expression for the bargaining weight of party 1 is obtained:
 
\begin{align}
\label{eq:perception}
\alpha &\equiv \frac{1}{2} \left[\alpha_1 + \alpha_2\right] \nonumber \\
    &=\frac{1}{2} + \frac{1}{4}\left[P_{1,1} + P_{1,2} -P_{2,1} - P_{2,2}\right] ~~~~  0 \leq P_{m,n} \leq 1
\end{align}

In the authors' previous work, based on perceptions, the bargaining weight demonstrated that parties' disagreement payoffs could impact their bargaining strength, illustrated as "Cases" in \citep{kryskowski2022applying}. This paper focuses on Cases 1 and 2, which depict scenarios where disagreement payoffs influence the bargaining weight. These Cases make the most straightforward assumptions about the negotiation process and provide valuable insights into cooperative and non-cooperative negotiation strategies.

\begin{table*}[]
\large
\caption{Symmetric Disagreement Payoff Driven Bargaining Weights}
\label{tab:symmetric}
\begin{tabular}{|c|c|c|c|c|c| }
\hline
\textbf{Case} & \textbf{P\textsubscript{1,1}} & \textbf{P\textsubscript{1,2}} & \textbf{P\textsubscript{2,1}} & \textbf{P\textsubscript{2,2}} & $\mathbf{\alpha}\left(d_1^\dagger,d_2^\dagger\right)$ \\ 
\hline
\textbf{1} & $d_1^\dagger$ & $d_1^\dagger$ & $ d_2^\dagger$ & $d_2^\dagger$ & $\frac{1}{2} + \frac{d_1^\dagger - d_2^\dagger}{2}$ \\
\hline
\textbf{2} & $\frac{d_1^\dagger}{d_1^\dagger + d_2^\dagger}$ & $\frac{d_1^\dagger}{d_1^\dagger+ d_2^\dagger}$ & $\frac{d_2^\dagger}{d_1^\dagger+ d_2^\dagger}$ & $\frac{d_2^\dagger}{d_1^\dagger + d_2^\dagger}$ &  $\frac{d_1^\dagger}{d_1^\dagger + d_2^\dagger}$\\
\hline
\end{tabular}
\end{table*}

\subsection{The Original Nash Bargaining Solution}

When $P_{1,1} + P_{1,2} = P_{2,1} + P_{2,2}$ in Eq. \eqref{eq:perception}, then  $\alpha=1/2$ and the Original/Classic symmetric NBS is obtained:

\begin{equation}
\label{eq:NBS}
\frac{r}{O_M}= d_1^\dagger + \frac{1}{2}\left(1 - d_2^\dagger -d_1^\dagger\right ) =\frac{1}{2} + \frac{d_1^\dagger - d_2^\dagger}{2}
\end{equation}

The Original NBS assumes a fixed bargaining weight of $\alpha =1/2$, representing equal bargaining power between the parties.

\subsection{Case 1}

In Case 1 of Table \ref{tab:symmetric}, each party perceives its bargaining strength as equal to its respective disagreement payoff. Furthermore, each party believes the other party's bargaining strength equals their respective disagreement payoff. Substituting Case 1 of Table \ref{tab:symmetric} into Eq. $\eqref{eq:NBTS6}$:

\begin{equation}
\label{eq:Case1}
\frac{r}{O_M}=  \frac{{d_2^\dagger}^2 - {d_1^\dagger}^2 + 2\left(d_1^\dagger-d_2^\dagger\right) + 1}{2} 
\end{equation}

Case 1 is notable for its cooperative and intuitive nature since each party perceives its own and the other party's strength as its respective disagreement payoff, a reasonable assumption. Consequently, the bargaining weight, $\mathbf{\alpha}\left(d_1^\dagger,d_2^\dagger\right)$, is equivalent to the Original NBS.

\subsection{Case 2}

Case 2 of Table \ref{tab:symmetric} represents a limiting case of the Rubinstein model \citep{rubinstein1982perfect}. Additionally, it is assumed that the parties' disagreement payoffs are a reasonable proxy for their patience \citep{muthoo1999bargaining}. Substituting Case 2 of Table \ref{tab:symmetric} into Eq. $\eqref{eq:NBTS6}$:

\begin{equation}
\label{eq:Case2}
\frac{r}{O_M}=\frac{d_1^\dagger}{d_1^\dagger+d_2^\dagger}= \frac{1}{1+ \frac{d_2^\dagger}{d_1^\dagger}}
\end{equation}

Case 2 is intriguing because each party's payoff corresponds to its bargaining weight. Moreover, the solution is independent of the operating income, making it a non-cooperative bargain.

\section{Establishing Disagreement Payoff Bounds}

Establishing accurate bounds for disagreement payoffs is crucial in applying the NBS to determine reasonable royalties, especially under the uncertainties typical in intellectual property disputes. Disagreement payoffs represent what each party expects to secure outside the negotiation context—effectively, their fallback positions if negotiations fail. This section outlines a robust methodology for defining these bounds, ensuring they reflect realistic economic potentials.

\subsection{Payoff Bounds}
\label{sec:Payoff Bounds}

In the context of intellectual property, determining precise disagreement payoffs poses substantial challenges due to the variable nature of market conditions and the proprietary aspects of technological innovations. These payoffs critically influence negotiation dynamics by setting the threshold below which parties are unlikely to settle, thus defining a bargaining range.

To address the uncertainty inherent in determining disagreement payoffs in the NBS, this paper proposes establishing bounds that encapsulate the potential range of outcomes. These bounds reflect each party's best-and-worst case scenarios based on available data and strategic considerations. The bounds for $d_1^\dagger$ and $d_2^\dagger$ are introduced to account for the uncertainty surrounding each party's own disagreement payoffs as well as the payoffs of the other party.

The disagreement payoffs are assumed to follow a uniform probability distribution between their respective limits, reflecting a lack of bias toward any particular outcome within the range. This assumption introduces randomness, accounting for the unpredictability inherent in real-world scenarios. Additionally, the disagreement payoffs are assumed to be statistically independent.

In practical terms, establishing these bounds allows negotiators to approach the bargaining table with a clearer understanding of their positions and limits. It also provides a framework to simulate various negotiation outcomes and better prepare for actual negotiations. Utilizing these bounds, negotiators can more accurately model the negotiation scenario using the NBS. This modeling helps forecast likely outcomes and prepares both parties for the most equitable and economically rational royalty arrangement.

The bounds on the disagreement payoffs are defined as follows:

\noindent
\begin{equation}
\label{eq:NBTS61}
0 \leq a \leq d_1^\dagger \leq b \leq 1
\end{equation}

\noindent

\begin{equation}
\label{eq:NBTS62}
0 \leq c \leq d_2^\dagger \leq d \leq 1
\end{equation}

\begin{equation}
\label{eq:Payoff3}
b + d \leq 1
\end{equation}

Given these ranges for the disagreement payoffs, estimating a reasonable royalty by applying the Bayes Cost Method \citep{melsa1978decision} is straightforward to account for uncertainty.

\section{Estimation Theory in a Two-Party Bargaining Context}
\label{sec:EstimationTheory}
This section explores the application of estimation theory in two-party bargaining scenarios, highlighting the role of cost functions in representing the economic risks tied to deviations from the actual royalty given a specific set of disagreement payoffs. These cost functions are crucial for determining the optimal royalty estimate, aligned with the parties' risk preferences, and play a pivotal role in the economic modeling of bargaining situations.

To realistically account for risk, this paper utilizes three common cost functions: Absolute-Value, Uniform, and Square Error, representing risk-neutral, risk-seeking, and risk-averse negotiations, respectively. These cost functions are integrated with the NBS using a Bayesian Cost approach for royalty estimation, which is discussed next.

\subsection{A Posteriori Density Functions}
\label{sec:posteriori}
Point estimates of the royalty involve the a posteriori probability density, $p(\boldsymbol{\theta} \, \vert \, \boldsymbol{d^\dagger})$, which describes the statistical uncertainty about the royalty (now considered a random variable) conditional on a collection of observed disagreement payoffs. The boldface symbols denote vector quantities, making the following equivalent:

\noindent
\begin{equation}
\label{eq:posteriori_1}
p(\boldsymbol{\theta} \, \vert \, \boldsymbol{d^\dagger}) = p(\{\theta_1, \theta_2 \} \, \vert \, \{ d_1^\dagger, d_2^\dagger\} )
\end{equation}

Only one royalty estimate, say $\hat{\theta}_1$,  is needed as $\hat{\theta}_2 \equiv 1 - \hat{\theta}_1$; therefore, for the rest of the paper, the a posteriori probability density:

\noindent
\begin{equation}
\label{eq:posteriori_2}
p(\theta_1 \, \vert \, \boldsymbol{d^\dagger}) = p(\theta \, \vert \, \boldsymbol{d^\dagger}) 
\end{equation}

will be used in the Bayes Cost Method to determine the optimal point estimates of the royalty.

Each $p(\theta \, \vert \, \boldsymbol{d^\dagger})$ for Case 1, Case 2, and the NBS is its corresponding functional form for royalty driven by the random variables $\{d_1^\dagger,d_2^\dagger\}$.

\subsection{Cost Functions}
\label{sec:DefiningCostFunctions}
A cost function, the negative of a utility function, encapsulates the risk inherent in the entire negotiation process, \cite{robert2007bayesian, geweke2005contemporary}. The cost function, denoted as $C(\boldsymbol{\theta},\boldsymbol{\hat{\theta}}(\boldsymbol{d^\dagger}))$, assigns to each combination of true royalty $\boldsymbol{\theta}= \{\theta_1, \theta_2\}$ and royalty estimate $\boldsymbol{\hat{\theta}}(\boldsymbol{d^\dagger}) = \{\hat{\theta}_1(d_1^\dagger,d_2^\dagger), \hat{\theta}_2(d_1^\dagger,d_2^\dagger)\}$ a unique cost.

It is assumed that both parties are adopting the same risk profile, so similar to the reasoning in Section \ref{sec:posteriori}, only party 1's royalty needs to be considered:

\begin{equation}
\label{eq:NBTS15}
C(\theta,\hat{\theta}(\boldsymbol{d^\dagger}))=C(\theta,\hat{\theta})=-U(\theta,\hat{\theta})
\end{equation}

In negotiation terms, parties aim to maximize their utility by minimizing the expected value of the cost, $E\left\{C(\theta,\hat{\theta})\right\}$. Cost functions capture the bargaining dynamics and provide a quantitative basis for resolving disputes by minimizing the expected "cost" or "loss" during negotiations. Each cost function listed below corresponds to a risk profile: risk-neutral, risk-seeking, or risk-averse, which influences the strategic behavior of the parties. These three cost functions are universal throughout estimation theory. For a detailed mathematical treatment of the Bayes Cost Method, see Melsa and Cohn \citep{melsa1978decision}.

\hspace{1cm}

1. \textbf{Absolute-Value Cost Function (ABS)}: The absolute value of the estimation error represents a risk-neutral stance and is defined by:

\begin{equation}
\label{eq:abs_1}
 C_\mathrm{ABS}(\theta,\hat{\theta})= |\theta-\hat{\theta}|
\end{equation}

The estimate $\hat{\theta}_{\mathrm{ABS}}$ is obtained by finding $\hat{\theta}$ that minimizes:

\begin{equation}
\label{eq:abs_2}
E\left\{C_\mathrm{ABS}(\theta,\hat{\theta})\right\} = \int_{0}^{1} |\theta-\hat{\theta}|\,p(\theta \, \vert \, \boldsymbol{d^\dagger}) \, d\theta
\end{equation}

The solution to Eq. \ref{eq:abs_2} is well known:  $\hat{\theta}_{\mathrm{ABS}}$ is the median of the a posteriori density $p(\theta \, \vert \, \boldsymbol{d^\dagger})$.  In other words, if $\mathrm{CDF}(\theta\,\vert\, \boldsymbol{d^\dagger})$ is the Cumulative Distribution Function of $\theta$ given $\boldsymbol{d^\dagger}$, then $\hat{\theta}_{\mathrm{ABS}}$ is defined as:

\begin{equation}
\label{eq:abs_3}
\mathrm{CDF}(\hat{\theta}_{\mathrm{ABS}} \,\vert\, \boldsymbol{d^\dagger}) = P\left\{ \theta \leq \hat{\theta}_{\mathrm{ABS}} \,\vert\, \boldsymbol{d^\dagger}\right\}= \frac{1}{2}
\end{equation}

Where $P\left\{ \theta \leq \hat{\theta}_{\mathrm{ABS}} \,\vert\, \boldsymbol{d^\dagger}\right\}$ is the probability that $\theta$ is less than or equal to $\hat{\theta}_\mathrm{ABS}$ given $\boldsymbol{d^\dagger}$.

\hspace{1cm}

2. \textbf{Uniform Cost Function (UC)}: Suitable for risk-seeking parties, this function is used when the objective is to achieve the most probable royalty. It is represented by:

\begin{equation}
\label{eq:UC_1}
C_\mathrm{UC}(\theta,\hat{\theta})= 
\begin{cases}
    0& \text{if }  |\theta-\hat{\theta}| < \epsilon \\
    1              & \text{otherwise}
\end{cases}
\end{equation}
The estimate  $\hat{\theta}_{\mathrm{UC}}$ is obtained by finding $\hat{\theta}$ that minimizes:

\begin{equation}
\label{eq:UC_2}
\begin{split}
 E\left\{C_\mathrm{UC}(\theta,\hat{\theta})\right\} & = \int_{|\theta-\hat{\theta}|\, > \, \epsilon }  \,p(\theta \, \vert \, \boldsymbol{d^\dagger}) \, d\theta \\
 & = \int_{0}^{1} p(\theta \, \vert \, \boldsymbol{d^\dagger}) \, d\theta -  \int_{\hat{\theta}-\epsilon}^{\hat{\theta}+\epsilon} p(\theta \, \vert \, \boldsymbol{d^\dagger}) \, d\theta  \\
 &= 1 -  \int_{\hat{\theta}-\epsilon}^{\hat{\theta}+\epsilon} p(\theta \, \vert \,\boldsymbol{d^\dagger}) \, d\theta  \\
 \end{split}
\end{equation}

To minimize $E\left\{C_\mathrm{UC}(\theta,\hat{\theta})\right\}$, $p(\hat{\theta} \, \vert \, \boldsymbol{d^\dagger})$ must be maximized. Therefore, the optimal estimator $\hat{\theta}_{\mathrm{UC}}$ is defined by:

\begin{equation}
\label{eq:UC_3}
p(\hat{\theta}_{\mathrm{UC}} \, \vert \, \boldsymbol{d^\dagger}) \geq p(\hat{\theta} \, \vert \, \boldsymbol{d^\dagger})
\end{equation}

for all $\hat{\theta} \neq \hat{\theta}_{\mathrm{UC}}$, where $\hat{\theta}_{\mathrm{UC}}$ represents the mode of the a posteriori density $p(\theta \, \vert \, \boldsymbol{d^\dagger})$. Since $\hat{\theta}_{\mathrm{UC}}$ maximizes the a posteriori density,  it is commonly referred to as the maximum a posteriori (MAP) estimator, denoted $\hat{\theta}_{\mathrm{MAP}}$.  This notation will be used in subsequent discussions due to its prevalence. However, it is important to note that since $p(\theta \, \vert \, \boldsymbol{d^\dagger})$ can be multimodal, the MAP estimator may not be unique. \\

\hspace{1cm}

3. \textbf{Square Error Cost Function (SE)}: This function appeals to risk-averse negotiators as it minimizes the variance of the outcomes, ensuring the least fluctuation from the true royalty. It is given by:

\begin{equation}
\label{eq:mse_1}
 C_\mathrm{SE}(\theta,\hat{\theta})= (\theta-\hat{\theta})^2 
\end{equation}

The estimate $\hat{\theta}_{\mathrm{SE}}$ is obtained by finding $\hat{\theta}$ that minimizes:

\begin{equation}
\label{eq:mse_2}
E\left\{C_\mathrm{SE}(\theta,\hat{\theta})\right\} = \int_{0}^{1} (\theta-\hat{\theta})^2\,p(\theta \, \vert \, \boldsymbol{d^\dagger}) \, d\theta
\end{equation}

The solution that minimizes Eq. \ref{eq:mse_2} is also well known and  $\hat{\theta}_{\mathrm{SE}}$ is the mean of the a posteriori density $p(\theta \, \vert \, \boldsymbol{d^\dagger})$.  This estimator is also known as the conditional mean estimator and is usually referred to as the Mean Square Error (MSE) and is given by:

\begin{equation}
\label{eq:mse_3}
\hat{\theta}_{\mathrm{MSE}} = \int_{0}^{1} \theta \, p(\theta \, \vert \, \boldsymbol{d^\dagger}) \, d\theta
\end{equation}

It is easily shown that $\hat{\theta}_{\mathrm{MSE}}$ is an unbiased estimate because the expected value of the estimation error is zero (i.e., $E\{\theta-\hat{\theta}\} = 0$). Moreover, the estimation error variance is the smallest of the three estimators.

\hspace{1cm}

\begin{table*}[]
\caption{MAP and ABS Royalty Estimate For Party 1}
\label{tab:symmetric1}
\begin{tabular}{|c|c|c| }
\hline
\textbf{Case} & $\hat{\theta}_\mathrm{MAP}$& $\hat{\theta}_\mathrm{ABS}$ \\
\hline
\textbf{NBS} & $b + 1/2\,(1-b-d)$& $1/4\,(a+b-c-d) + 1/2$  \\
\hline
\textbf{1} & $1/2\,(d^2-b^2+1) + b - d$ &$1/8\,(-{a}^{2}+\left( -2\,b+4 \right) a-{b}^{2}+
 \left( c+d-2 \right) ^{2}) + b/2$ \\
\hline
\textbf{2} & $b/(b+d)$ & $(a+b)/(a+b+c+d)$ \\
\hline
\end{tabular}
\end{table*}

\begin{table*}[]
\caption{MSE Royalty Estimate For Party 1}
\label{tab:symmetric2}
\begin{tabular}{|c|c| }
\hline
\textbf{Case} & $\hat{\theta}_\mathrm{MSE}$ \\
\hline
\textbf{NBS} & $1/4\,(a+b-c-d) + 1/2$ \\
\hline
\textbf{1} & $ 1/6\,({c}^{2}+d\,c+{d}^{2}-{a}^{2}-a\,b-{b}^{2}) + 1/2\,(a+b-c-d+1)$ \\
\hline
\textbf{2} & \large{$ {\frac { \left( {a}^{2}-{c}^{2} \right) \ln  \left( a+c \right) + 
\left( {d}^{2} - a^{2}\right) \ln  \left( a+d \right) + \left( {c}^{2} - b^{2} \right) \ln  \left( b+c \right) + \left( {b}^{2}-{d}^{2} 
 \right) \ln  \left( b+d \right) + \left( c-d \right)  \left( a-b 
 \right) }{2\, \left( c-d \right)  \left( a-b \right) }}$} \\
\hline
\end{tabular}
\end{table*}

\subsection{Visualizing Cost Functions}

\begin{figure}[]
\begin{center}
\includegraphics[scale=.35]{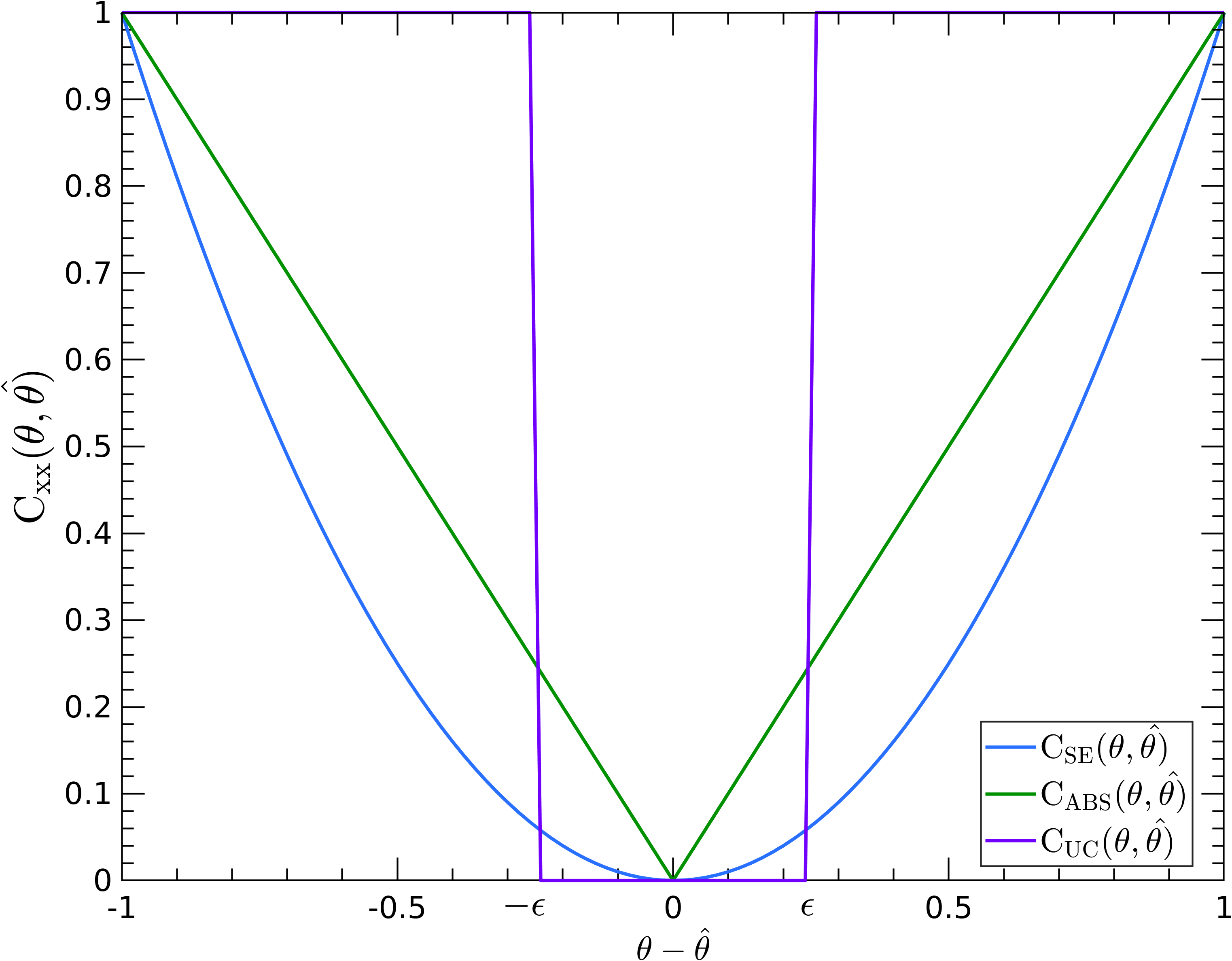} 
\end{center}
\caption{Common Cost Functions}
\label{fig:Cost}
\end{figure}

Fig. \ref{fig:Cost} offers a clear view of the behavior of each cost function. The x-axis represents the difference between the true royalty and its estimate, and the y-axis represents the corresponding cost or dissatisfaction when the estimate differs from the true royalty. 

The ABS cost function, depicted in green, increases linearly, reflecting a constant dissatisfaction rate across all deviations. The purple line represents the UC cost function, indicating no penalty within a tolerance range $-\epsilon \, \leq \theta-\hat{\theta} \, \leq +\epsilon$, and a unit penalty outside the tolerance range. The blue line represents the SE cost function, which exhibits a quadratic growth in cost, penalizing deviations more severely as they grow. These visualizations not only aid in understanding the quantitative aspects of cost functions but also emphasize their strategic implications in royalty negotiations, underscoring the importance of selecting a cost function that aligns with the negotiators' risk preferences.


\section{Determining the Royalty Estimate}
\label{sec:RoyaltyEstimate}
This section showcases the computation of the royalty estimate for each cost function as applied to the Cases and NBS. In addition, this section offers user-friendly tables for computing the royalty estimates. Table \ref{tab:symmetric1} provides the ABS and MAP royalty estimates, while Table \ref{tab:symmetric2} provides the MSE royalty estimate. To use these tables, substitute the numerical disagreement payoff bounds into the table(s) based on the Case and risk preference.
 
\subsection{Determining the ABS Royalty Estimate}

The ABS estimate of the royalty, $\hat{\theta}_{\mathrm{ABS}}$, is determined as the median of the a posteriori density $p(\theta \, \vert \, \boldsymbol{d^\dagger})$. The estimate $\hat{\theta}_{\mathrm{ABS}}$ in Table \ref{tab:symmetric1} is derived from the expected value of $d_1^\dagger$ and $d_2^\dagger$:

\begin{equation}
\label{eq:Ed1}
E\left\{d_1^\dagger\right\}= \frac{b+a}{2} 
\end{equation}

\begin{equation}
\label{eq:Ed2}
E\left\{d_2^\dagger\right\}= \frac{d+c}{2} 
\end{equation}

To determine $\hat{\theta}_{\mathrm{ABS}}$, substitute $\boldsymbol{d^\dagger} = \left\{E\left\{d_1^\dagger\right\},E\left\{d_2^\dagger\right\}\right\}$ into the NBS (Eq. \ref{eq:NBS}) and Cases (Eqs \ref{eq:Case1} - \ref{eq:Case2}). The results of these substitutions are presented in Table \ref{tab:symmetric1}. However, for Case 1, this method is only an approximation where generally  $P\left\{ \theta \leq \hat{\theta}_{\mathrm{ABS}} \,\vert\, \boldsymbol{d^\dagger}\right\} \neq \frac{1}{2}$. Despite this, the approximation of $\hat{\theta}_{\mathrm{ABS}}$ is remarkably accurate to within $\pm 4\%$ from its true value. 

$\hat{\theta}_{\mathrm{ABS}}$ is particularly useful because each party's royalty is adjusted to maintain a 50/50 probability of underpayment or overpayment. $\hat{\theta}_{\mathrm{ABS}}$ should be chosen when the parties aim for a fair, median-based approach without disproportionately favoring either side.

\subsection{Determining The MAP Royalty Estimate}

The MAP estimate, $\hat{\theta}_{\mathrm{MAP}}$, corresponds to the conditional mode of the a posteriori density function--that is, the royalty with the highest likelihood of occurring given $\boldsymbol{d^\dagger} $ as indicated by  Eq. \ref{eq:UC_3}. To find $\hat{\theta}_{\mathrm{MAP}}$, substitute $\boldsymbol{d^\dagger} = \{b,d\}$ into the NBS and Cases.  The results of these substitutions are found in Table \ref{tab:symmetric1}.

One rationale for choosing $\hat{\theta}_{\mathrm{MAP}}$ is the belief that if an agreement is not reached, each party will maximize their earning potential from outside options. This represents risk-seeking behavior, as parties aim to maintain their best negotiating position. $\hat{\theta}_{\mathrm{MAP}}$ is ideal for scenarios where achieving the most likely outcome is prioritized over mitigating risk. In summary, $\hat{\theta}_{\mathrm{MAP}}$ is critical for decision-making in contexts where the most likely estimate is required.

\subsection{Determining the MSE Royalty Estimate}

The MSE estimate of the royalty, $\hat{\theta}_{\mathrm{MSE}}$, corresponds to the conditional mean of the a posteriori density function. To find $\hat{\theta}_{\mathrm{MSE}}$, the expected value of $p(\theta \, \vert \, \boldsymbol{d^\dagger})$ must be performed. For the NBS, finding this expected value is straightforward and is identical to $\hat{\theta}_{\mathrm{ABS}}$. For Case 1, $\hat{\theta}_{\mathrm{MSE}}$ is given by:

\begin{equation}
\label{eq:ECase1}
\hat{\theta}_{\mathrm{MSE}} =  \frac{E\left\{{d_2^\dagger}^2\right\} - E\left\{{d_1^\dagger}^2\right\} + 2\left(E\left\{d_1^\dagger\right\}-E\left\{d_2^\dagger\right\}\right) + 1}{2} 
\end{equation}

The expected values of ${d_1^\dagger}^2$ and ${d_2^\dagger}^2$ are formed using the variances of $d_1^\dagger$ and $d_2^\dagger$: 

\begin{equation}
\label{eq:Vd1}
Var\left\{d_1^\dagger\right\}= \frac{(b-a)^2}{12} 
\end{equation}

\begin{equation}
\label{eq:Vd2}
Var\left\{d_2^\dagger\right\}= \frac{(d-c)^2}{12} 
\end{equation}

Thus, the expected values are:

\begin{equation}
\label{eq:Ed1sq}
E\left\{{d_1^\dagger}^2\right\} = Var\left\{d_1^\dagger\right\} + E\left\{d_1^\dagger\right\}^2
\end{equation}

\begin{equation}
\label{eq:Ed2sq}
E\left\{{d_2^\dagger}^2\right\} = Var\left\{d_2^\dagger\right\} + E\left\{d_2^\dagger\right\}^2
\end{equation}

Using Eqs. \ref{eq:Ed1} - \ref{eq:Ed2} and \ref{eq:Vd1} - \ref{eq:Ed2sq}, the $\hat{\theta}_{\mathrm{MSE}}$ can be obtained from Eq. \ref{eq:ECase1}.

For Case 2, obtaining $\hat{\theta}_{\mathrm{MSE}}$ is more complex, requiring the use of the software package Maple\footnote{Maple is a trademark of Waterloo Maple Inc.} for computation. The results are presented in Table \ref{tab:symmetric2}.

The economic rationale for choosing $\hat{\theta}_{\mathrm{MSE}}$ lies in the preference for risk aversion during deal negotiations. Therefore, $\hat{\theta}_{\mathrm{MSE}}$ is optimal for parties looking to reduce risk by minimizing the variance of the estimated royalty.

\begin{table*}[!ht]
\caption{Optimal Royalty Estimate For Party 1 where $a=0.00$, $b=0.20$, $c=0.00$, $d=0.80$}
\label{tab:data1}
\begin{tabular}{|c|c|c|c|c|c|c| }
\hline
\textbf{Case}  & $\hat{\theta}_\mathrm{MAP}$& $\hat{\theta}_\mathrm{ABS}$ & $\hat{\theta}_\mathrm{MSE}$ & $P\left\{ \theta \leq \hat{\theta}_\mathrm{MAP} \,\vert\, \boldsymbol{d^\dagger}\right\}$ &  $P\left\{ \theta \leq \hat{\theta}_\mathrm{ABS} \,\vert\, \boldsymbol{d^\dagger}\right\}$ &$P\left\{ \theta \leq \hat{\theta}_\mathrm{MSE} \,\vert\, \boldsymbol{d^\dagger}\right\}$ \\
\hline
\textbf{NBS} & 0.200 & 0.350 & 0.350 & 0.125 & 0.500 & 0.500 \\
\hline
\textbf{1}  & 0.200 & 0.275 & 0.300 & 0.308 & 0.495 & 0.547 \\
\hline
\textbf{2}  & 0.200 & 0.200 & 0.255 & 0.500 & 0.500 & 0.635 \\
\hline
\end{tabular}
\end{table*}

\section{Royalty Estimation Example}

This section provides an example of how royalty estimates are obtained by applying various cost functions to the Cases and NBS. This example demonstrates the practical application of the theoretical principles discussed in Sections \ref{sec:EstimationTheory} - \ref{sec:RoyaltyEstimate} and highlights the nuances of using different cost functions under varied circumstances. 

\begin{figure*}[hbt!]

\begin{subfigure}{.49\linewidth}
  \includegraphics[width=\linewidth]{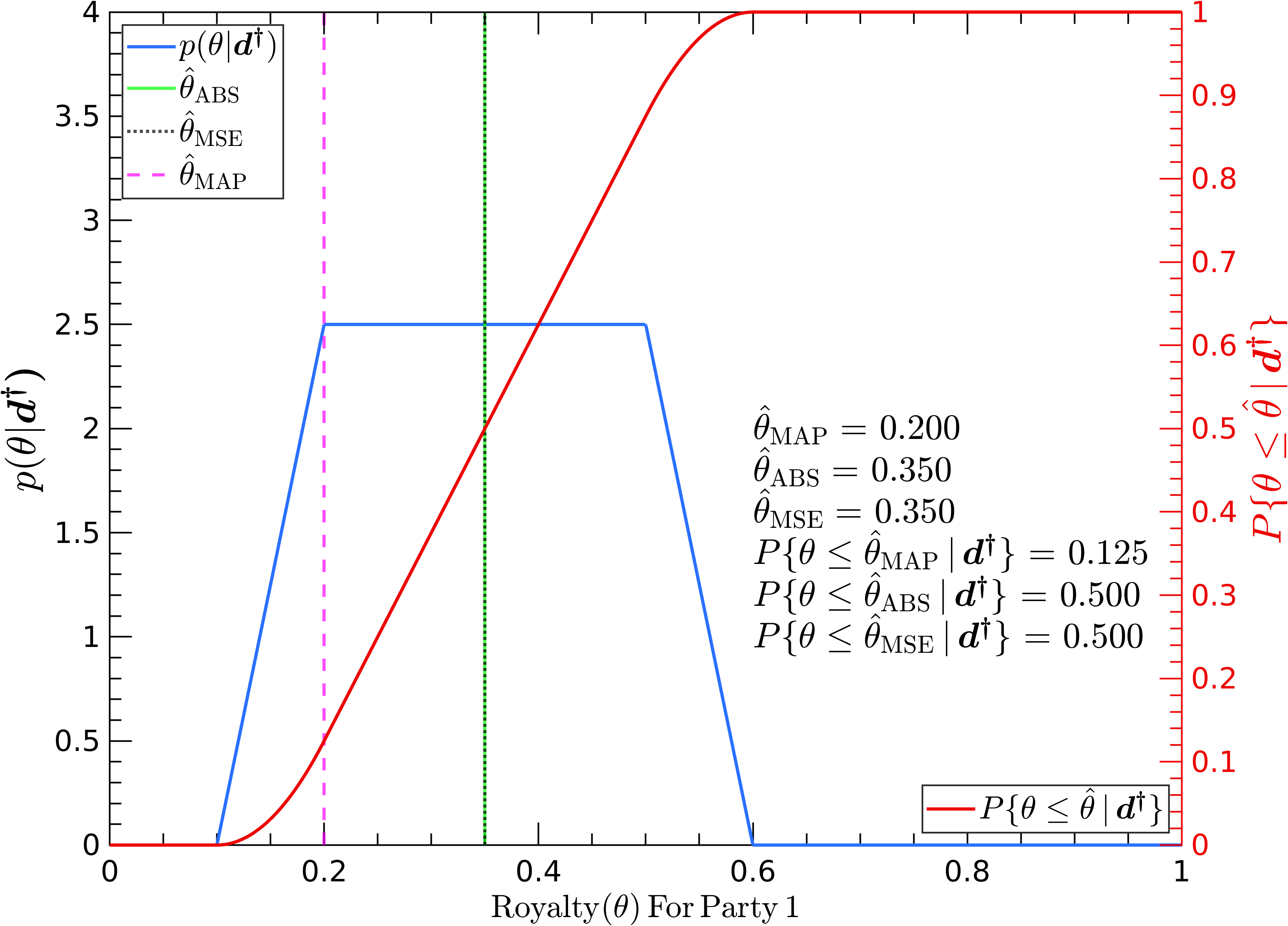}
  \caption{NBS : $d_1^{\dagger} = $ Uniform(0,0.2) : $d_2^{\dagger}$ = Uniform(0,0.8)}
  \label{fig:C0E0P1}
\end{subfigure}
\begin{subfigure}{.49\linewidth}
  \includegraphics[width=\linewidth]{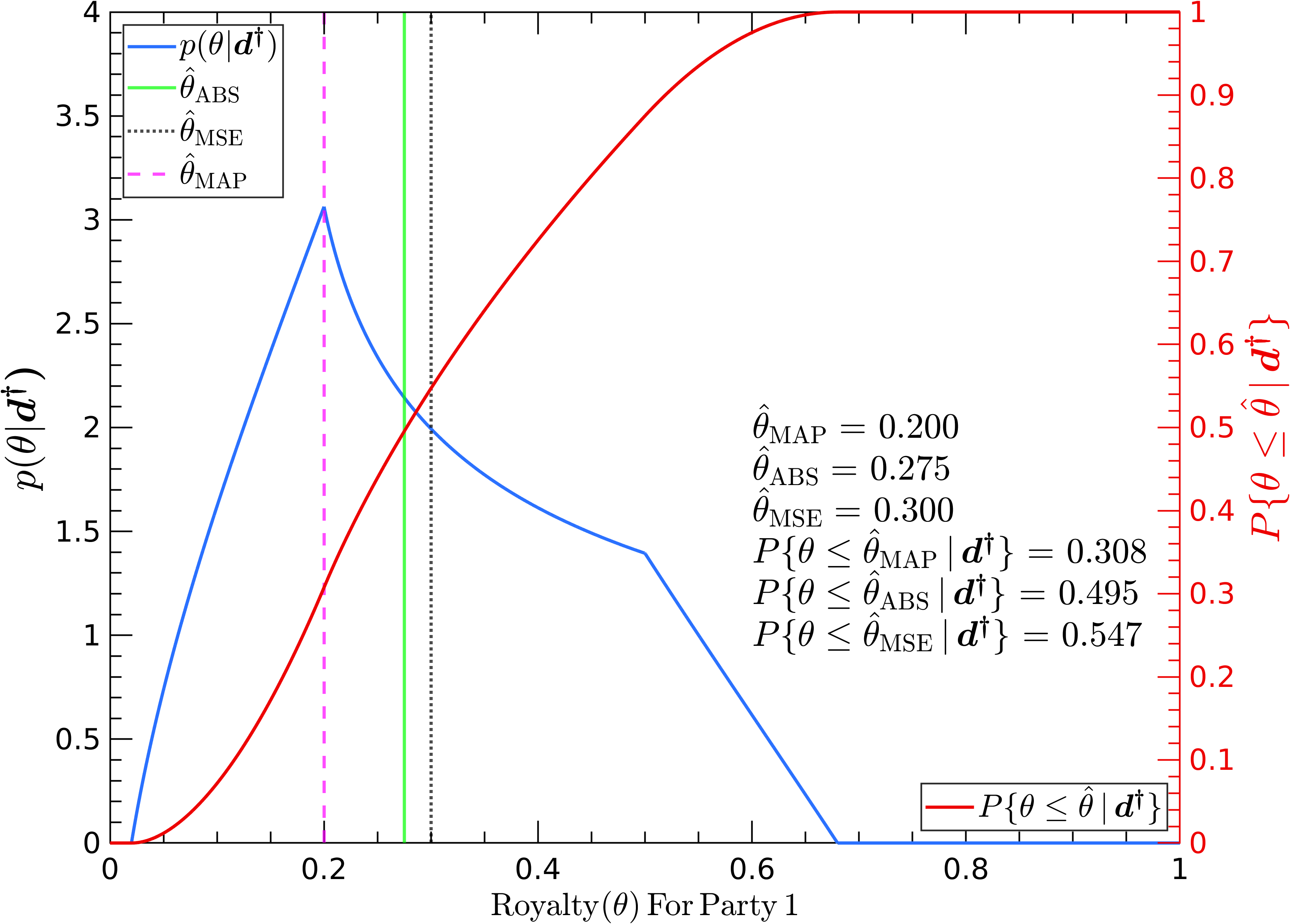}
  \caption{Case 1 : $d_1^{\dagger} = $ Uniform(0,0.2) : $d_2^{\dagger}$ = Uniform(0,0.8)}
  \label{fig:C1E0P1}
\end{subfigure}\hfill 

\medskip 

\begin{subfigure}{.49\linewidth}
  \includegraphics[width=\linewidth]{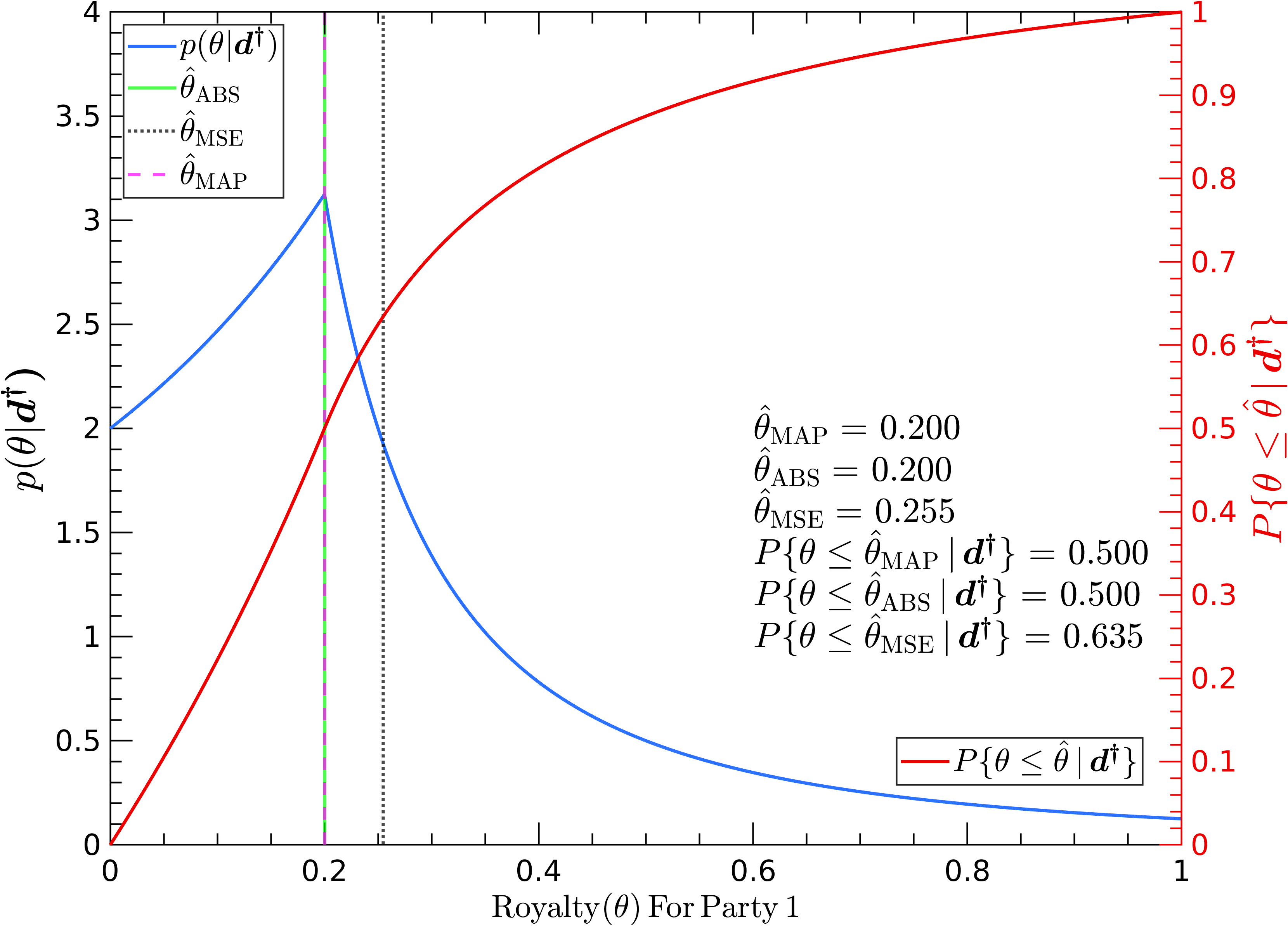}
  \caption{Case 2 : $d_1^{\dagger} = $ Uniform(0,0.2) : $d_2^{\dagger}$ = Uniform(0,0.8)}
  \label{fig:C2E0P1}
\end{subfigure}
\begin{subfigure}{.49\linewidth}
  \includegraphics[width=\linewidth]{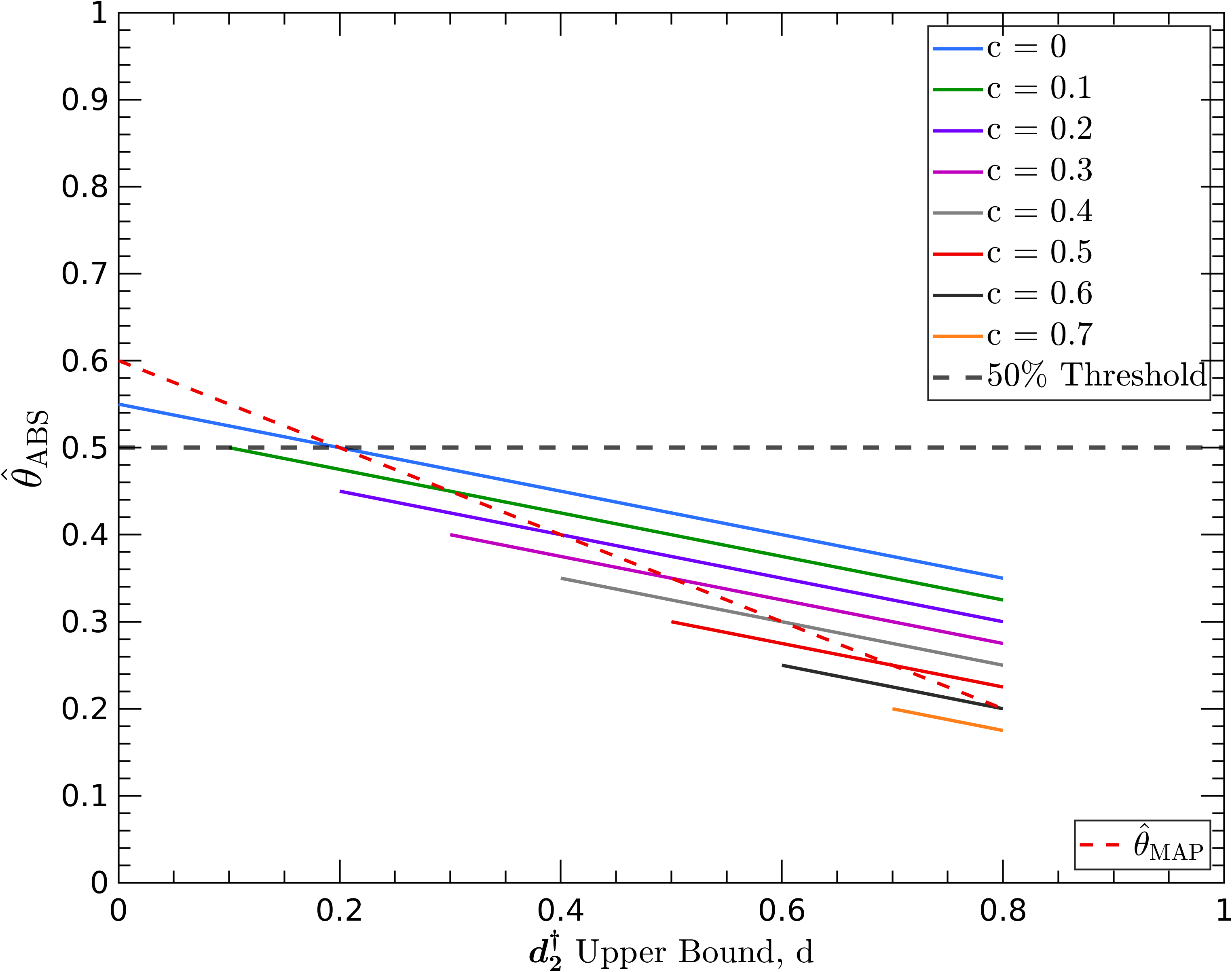}
  \caption{Family of $\hat{\theta}_{\mathrm{ABS}} = \hat{\theta}_{\mathrm{MSE}}$ for NBS}
  \label{fig:NBS_ABS}
\end{subfigure}\hfill 

\caption{}
\label{fig:FirstFour}
\end{figure*}

The hypothetical bounds are given in Table \ref{tab:data1}, where $a=0.00$, $b=0.20$, $c=0.00$, and $d=0.80$. The solutions are presented from the perspective of party 1.  The a posteriori densities presented in Figs. \ref{fig:C0E0P1} - \ref{fig:C2E0P1} were derived analytically using Maple; however, they can also be generated using Monte Carlo procedures. 

In Figs.  \ref{fig:C0E0P1} - \ref{fig:C2E0P1}, the x-axis represents the values for the royalty, $\theta$. The left y-axis represents $p(\theta \,  \vert \, \boldsymbol{d^\dagger})$, the a posteriori density, while the right y-axis represents the Cumulative Distribution Function, $\mathrm{CDF}(\hat{\theta} \,\vert\, \boldsymbol{d^\dagger}) = P\left\{ \theta \leq \hat{\theta} \,\vert\, \boldsymbol{d^\dagger}\right\}$, which shows the probability of royalty overpayment.

\subsection{Original NBS--Party 1's Perspective}

Fig. \ref{fig:C0E0P1} illustrates the a posteriori density for the Original NBS and royalty estimates using the three different cost functions. For the Original NBS, the a posteriori density generally forms a trapezoidal shape, implying the a posteriori density is multimodal. The conditional mode, $\hat{\theta}_\mathrm{MAP}$, does not necessarily equal the conditional mean and conditional median. Due to the symmetry of the conditional density $p(\theta \, \vert \, \boldsymbol{d^\dagger})$,  $\hat{\theta}_\mathrm{MSE}$ = $\hat{\theta}_\mathrm{ABS}$. 

However, if the conditional density is both symmetric and unimodal—implying that the span of $d_1^\dagger$ and $d_2^\dagger$ (the width of the distributions) are identical—then $\hat{\theta}_\mathrm{MAP}=\hat{\theta}_\mathrm{ABS}=\hat{\theta}_\mathrm{MSE}$. In such symmetric and unimodal scenarios, the a posteriori density forms a triangular shape, and if the $d_1^\dagger$ and $d_2^\dagger$ distributions are identical, then for all cases, $\hat{\theta}_\mathrm{MAP}=\hat{\theta}_\mathrm{ABS}=\hat{\theta}_\mathrm{MSE}=1/2$ for the royalty. 

In Fig. \ref{fig:C0E0P1}, $\hat{\theta}_\mathrm{MAP}$ is shown with a magenta dashed vertical line at $\hat{\theta}_\mathrm{MAP}=0.200$ with the probability $P\left\{ \theta \leq \hat{\theta}_\mathrm{MAP} \,\vert\, \boldsymbol{d^\dagger}\right\}=0.125$ indicating a 12.5\% chance that party 1 was overpaid. It should be noted that because the a posteriori density is flat in the range $0.200\leq\theta\leq0.500$, the mode is equally likely in this range. However, the point estimate for $\hat{\theta}_\mathrm{MAP}$ is chosen at the corner of the trapezoid due to its reliance solely on the disagreement payoff maximums. This results in a significantly lower royalty because party 1 has much less earning potential than party 2 if no deal is struck. Because the upper bounds for $d_1^\dagger$ and $d_2^\dagger$ add to unity, $\hat{\theta}_\mathrm{MAP}$ is the maximum disagreement payoff of party 1.

The green vertical line forms $\hat{\theta}_\mathrm{ABS}$. The probability $P\left\{ \theta \leq \hat{\theta}_\mathrm{ABS} \,\vert\, \boldsymbol{d^\dagger}\right\}=\frac{1}{2}$, indicates that there is a 50\% probability that the true value, ${\theta_\mathrm{ABS}}$, is  $\leq$ 0.350. This makes ${\hat{\theta}_\mathrm{ABS}}$ a median estimator, marking the value below and above which the cumulative probability is equally distributed. In simpler terms, half of the potential royalty values fall below this estimate, and half exceed it, positioning $\hat{\theta}_\mathrm{ABS}$ as the central value in the distribution.

The black dashed vertical line indicates $\hat{\theta}_\mathrm{MSE}$, which is unbiased with minimum variance, a key aspect in choosing a reliable estimator in statistical modeling.

\subsection{Case 1--Party 1's Perspective}

Fig. \ref{fig:C1E0P1} presents Case 1 and shows what happens when the NBS is substituted for the bargaining weight ($\alpha$). In this scenario, the a posteriori density clearly illustrates that $\hat{\theta}_\mathrm{MAP}$ occurs at the mode of the density function, $\hat{\theta}_\mathrm{ABS}$ sits at the median, and $\hat{\theta}_\mathrm{MSE}$ occurs at the mean.

$\hat{\theta}_\mathrm{MAP}$ represents the mode, which is the value ${\theta}$ that maximizes $p(\theta \, \vert \, \boldsymbol{d^\dagger})$. In other words, it is the most likely value of the royalty for party 1. The placement of $\hat{\theta}_\mathrm{MAP}$ at 0.200 clearly shows that this value of ${\theta}$ has the highest likelihood of being the true parameter value based on $p(\theta \, \vert \, \boldsymbol{d^\dagger})$.

The implications of these findings are significant. In this case, $\hat{\theta}_\mathrm{MSE}$ produces the most substantial royalty for party 1, while $\hat{\theta}_\mathrm{MAP}$ produces the lowest. $\hat{\theta}_\mathrm{MAP}$ results in a royalty of 0.200, $\hat{\theta}_\mathrm{ABS}$ produces a royalty of 0.275 (a more accurate value is 0.277 found by numerically solving Eq. \ref{eq:abs_3} for $\hat{\theta}_\mathrm{ABS}$), and $\hat{\theta}_\mathrm{MSE}$ falls at 0.300. Notice $\hat{\theta}_\mathrm{MSE}$ is typically very close to $\hat{\theta}_\mathrm{ABS}$.  

\subsection{Case 2--Party 1's Perspective}

Fig. \ref{fig:C2E0P1} displays Case 2, which is the limiting case of the Rubinstein model. The a posteriori density peaks at ${\theta}=0.200$ and decreases sharply, suggesting a skew toward lower ${\theta}$ values.

Fig. \ref{fig:C2E0P1} shows that $\hat{\theta}_\mathrm{ABS}$ produces the same royalty as the $\hat{\theta}_\mathrm{MAP}$, both yielding a royalty of 0.200.  The fact that $\hat{\theta}_\mathrm{MAP}$ and $\hat{\theta}_\mathrm{ABS}$ coincide suggests that the most likely value (mode) is also considered a central estimate in this model, a less common occurrence unless the distribution is symmetric or specially configured to reflect such properties.

The placement of $\hat{\theta}_\mathrm{MSE}$ at 0.255 suggests it offers a balance between minimizing errors and capturing the central tendency of the distribution because it is positioned away from the peak where sensitivity to small changes in $\theta$ are large.

\section{Family of Solutions for the NBS and Cases}

\begin{figure*}[hbt!]

\begin{subfigure}{.49\linewidth}
    \includegraphics[width=\linewidth]{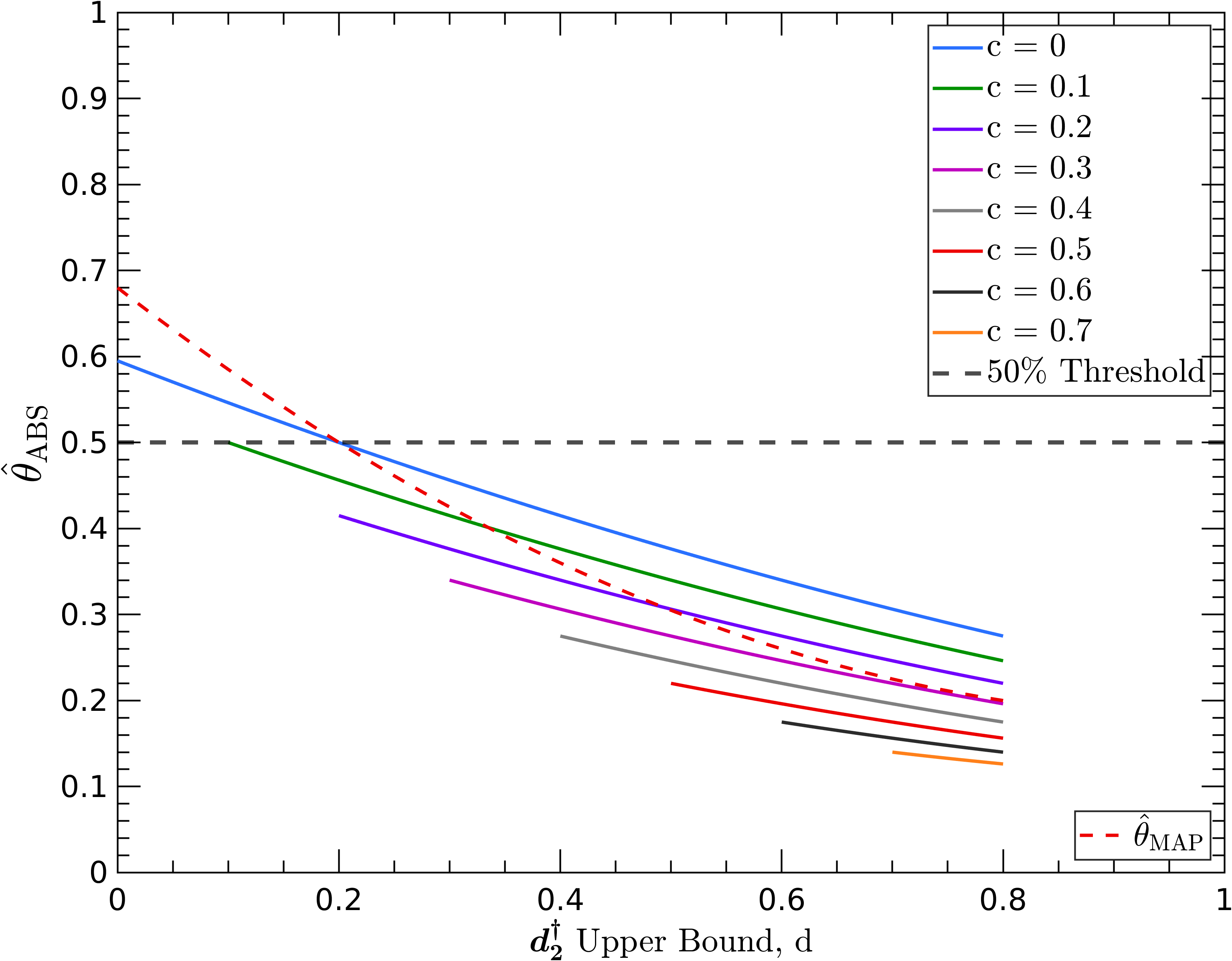}
  \caption{Family of $\hat{\theta}_{\mathrm{ABS}}$ for Case 1}
\label{fig:C1_ABS}
\end{subfigure}
\begin{subfigure}{.49\linewidth}
  \includegraphics[width=\linewidth]{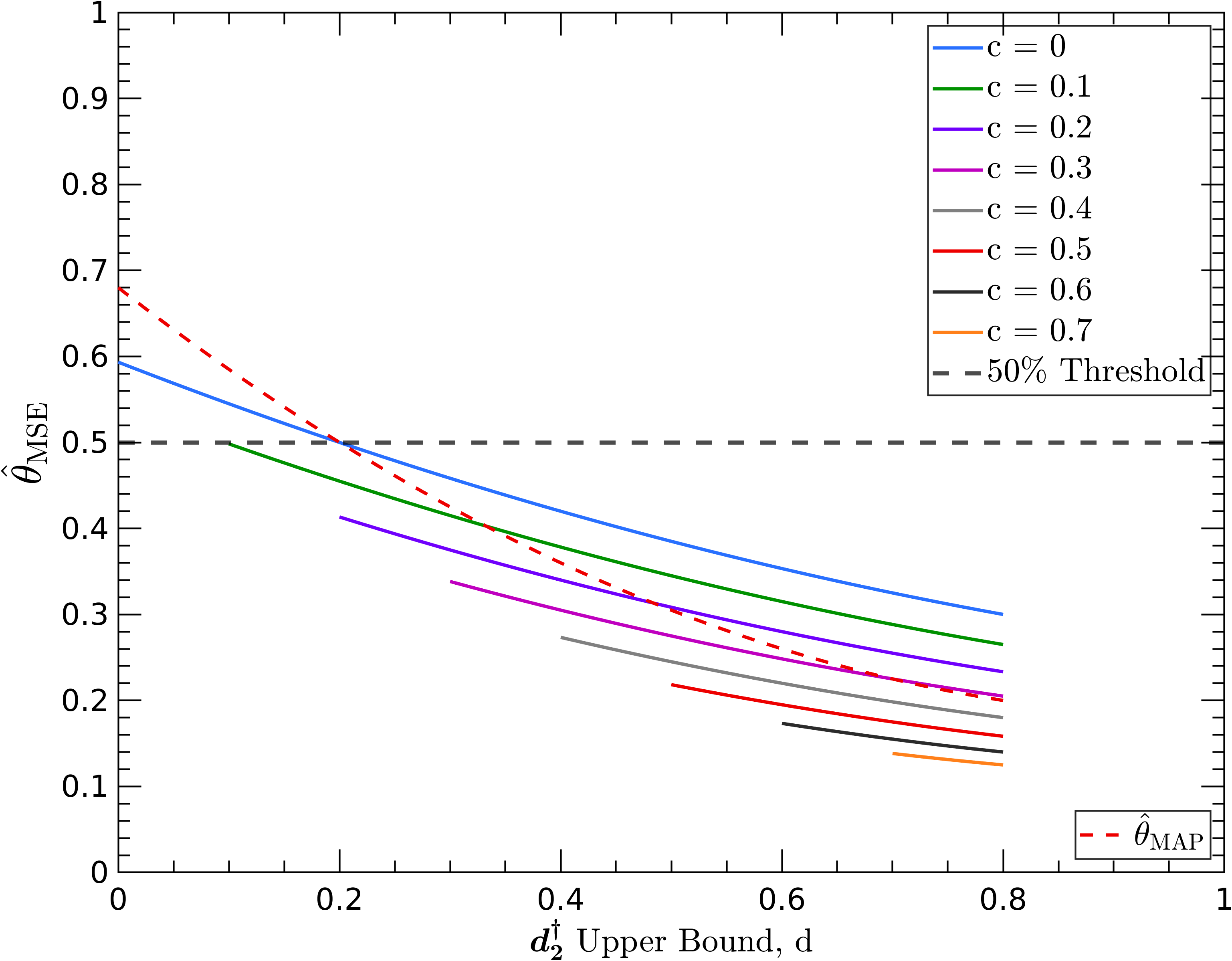}
  \caption{Family of $\hat{\theta}_{\mathrm{MSE}}$ for Case 1}
  \label{fig:C1_MSE}
\end{subfigure}\hfill 

\medskip 

\begin{subfigure}{.49\linewidth}
  \includegraphics[width=\linewidth]{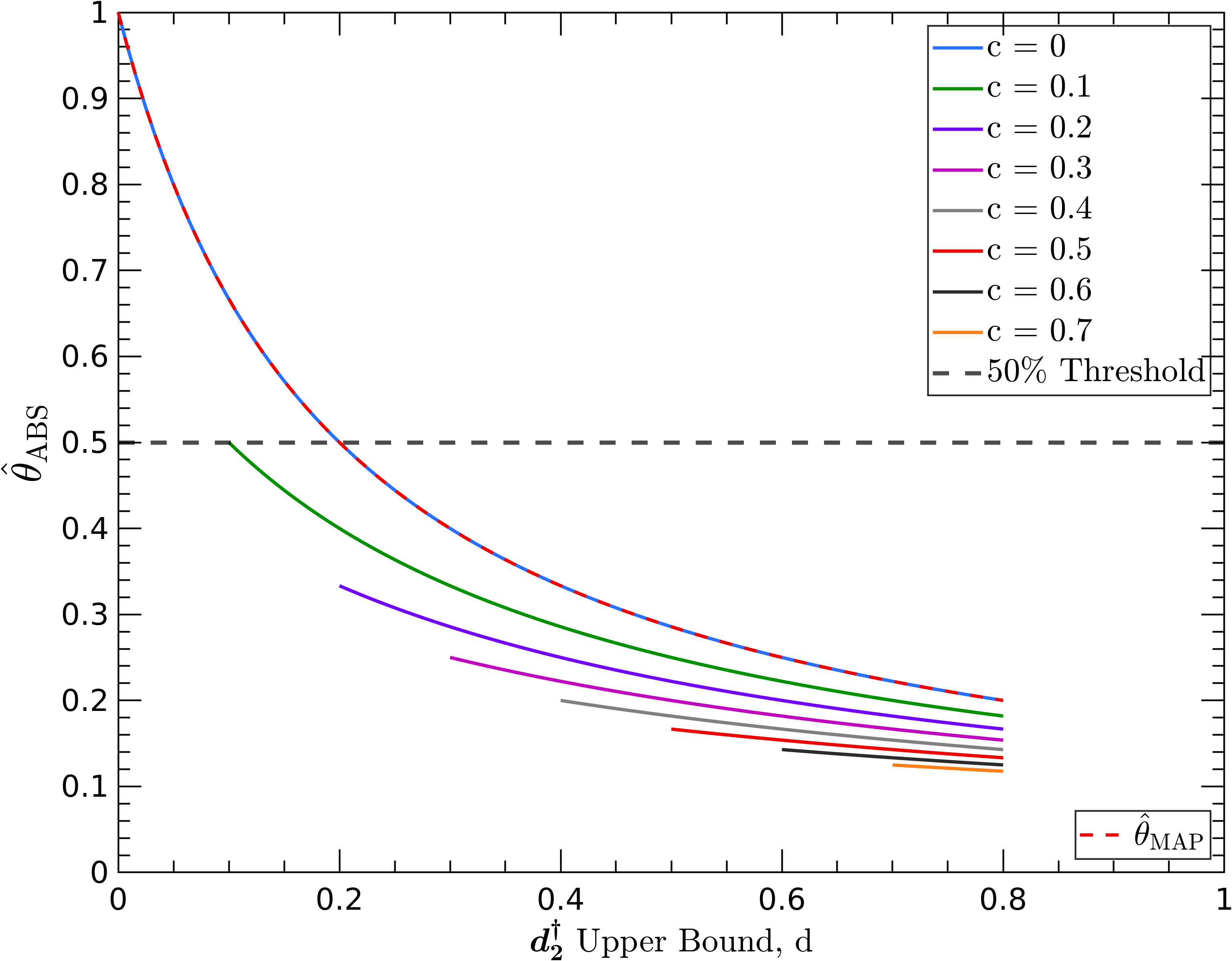}
  \caption{Family of $\hat{\theta}_{\mathrm{ABS}}$ for Case 2}
  \label{fig:C2_ABS}
\end{subfigure}
\begin{subfigure}{.49\linewidth}
  \includegraphics[width=\linewidth]{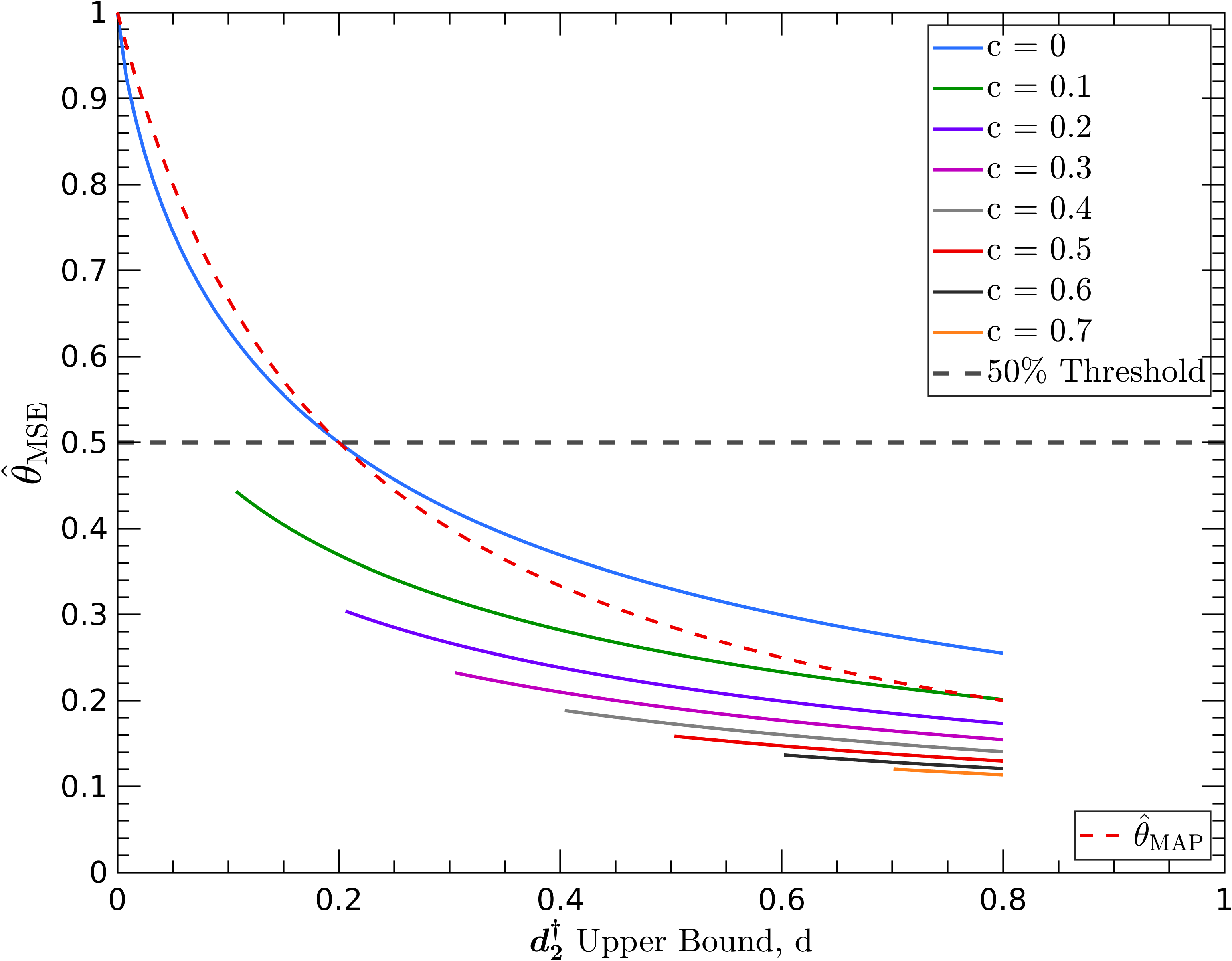}
  \caption{Family of $\hat{\theta}_{\mathrm{MSE}}$ for Case 2}
  \label{fig:C2_MSE}
\end{subfigure}\hfill 

\caption{}
\label{fig:SecondFour}
\end{figure*}

The next section presents a family of royalty estimates by fixing party 1's bounds at  $a=0.00$ and $b=0.20$ as shown in Fig. \ref{fig:NBS_ABS} and Figs. \ref{fig:C1_ABS} - \ref{fig:C2_MSE}. The x-axis represents the upper bound of party 2's disagreement payoff, $d$, and the individual lines are driven by party 2's lower bound, $c$. The y-axis represents either $\hat{\theta}_{\mathrm{ABS}}$ or $\hat{\theta}_{\mathrm{MSE}}$ with $\hat{\theta}_{\mathrm{MAP}}$ depicted as a red dashed line for visual reference.

These figures demonstrate the impact of each cost function on the three bargaining models. Comparing these figures reveals that certain bargaining models and cost functions benefit one party more than the other.

\subsection{Comparing the Family of Solutions}

Fig. \ref{fig:NBS_ABS} and Figs. \ref{fig:C1_ABS} - \ref{fig:C2_MSE}  illustrate the contrasts between the NBS and Cases for each cost function. The general conclusion is that the MSE favors the weaker party, the MAP favors the stronger party, and the ABS offers a neutral stance. Similarly, the NBS favors the weaker party, Case 2 favors the stronger party, and Case 1 provides a neutral stance.

The Original NBS in Fig. \ref{fig:NBS_ABS} lacks the flexibility to accommodate unequal bargaining power, ultimately tilting the balance in favor of the weaker party.

For Case 1, Figs. \ref{fig:C1_ABS} - \ref{fig:C1_MSE} show how profit shifts from one party to the other based on bargaining strength. The shift is mild, indicating a cooperative negotiation. Even if party 2's position is $\{c=0 \ , \, d=0\}$--that is, party 2 has no outside alternative--party 2's worth in the negotiation is still recognized.

Figs. \ref{fig:C2_ABS} - \ref{fig:C2_MSE} highlight Case 2, the Rubinstein model, where substantial penalties are imposed on the weaker party. The non-cooperative nature of this bargaining model is evident when party 2's position is $\{c=0 \ , \, d=0\}$, as party 1 will take all the profit and not recognize party 2's worth.

$\hat{\theta}_{\mathrm{ABS}}$ plays a crucial role, providing a fair standard with a 50/50 baseline for overpayment or underpayment by the parties. This function helps maintain a balanced negotiation. Weaker parties often have strong incentives to argue for a royalty that aligns with $\hat{\theta}_{\mathrm{MSE}}$, while stronger parties may prefer $\hat{\theta}_{\mathrm{MAP}}$.

\section{A Reasonable Royalty}

The following section explores the question: What is a reasonable royalty? From the perspective of \emph{Georgia Pacific} fifteen, this requires a bargaining model, and, as discussed in Section \ref{sec:models}, there is more than one model to choose from. Moreover, Section \ref{sec:EstimationTheory} discussed the concepts of risk and uncertainty and how they affect royalty determination by introducing three risk models that can be applied to each bargaining model. 

By comparing and contrasting these models, the authors conclude that Case 1, supported by the ABS cost function, provides the most reliable framework for determining a reasonable royalty.

\subsection{Reasonable Bargaining Model}

Case 1 should form the basis for a reasonable royalty because it assumes that each party perceives its own and the other's strengths as equivalent to their disagreement payoffs, thus equating the bargaining weight to the NBS. This assumption is highly rational, as the parties' perceived strengths are their opportunity costs. Case 1 is particularly valuable as it presumes parties negotiate in good faith, considering their operating income. This approach ensures cooperation and fairness with straightforward assumptions and aligns well with the implied good faith criteria found in \emph{Georgia Pacific} factor fifteen.

The Original NBS, however, does not align with human intuition because it assumes equal bargaining power for both parties, resulting in an even split of the surplus. This assumption is often unrealistic because parties frequently have unequal bargaining power. While it assumes cooperation, it is too rigid in its presumption of a 50/50 split of the surplus. Consequently, parties with poor opportunity costs might advocate for this model, as the equal split could result in a substantial return despite their weak disagreement payoff position.

On the other hand, in Case 2, the Rubinstein model represents a non-cooperative negotiation process. Here, parties do not consider their operating income or profit during negotiations. This model violates the main principles behind \emph{Georgia Pacific} factor fifteen, which emphasizes the desire to do business. Additionally, the party with a significantly better opportunity cost potential will likely recommend this model, which can yield substantially higher profits due to the strong bias against the weaker party.

\subsection{Reasonable Risk Model}

In the authors' view, the ABS estimator is the most fair and equitable because $\hat{\theta}_{\mathrm{ABS}}$ does not favor one side over the other, providing an equal 50/50 probability of overpayment or underpayment for both parties, thus treating each party equally. The parties consider the entire range of opportunity costs, and averaging this range is a reasonable way to gauge strength, which is a highly rational assumption. From human experience, when parties negotiate, they do not want to feel cheated, so they often aim for a fair deal that satisfies both parties, facilitating future business interactions.

On the other hand, the MAP estimator, designed to produce the most probable royalty, is best suited for situations where the parties are eager to reach a deal. This mindset often stems from a sense of urgency, leading to a rush to judgment that the other party will achieve their highest disagreement payoff. However, $\hat{\theta}_{\mathrm{MAP}}$ favors the party with the larger disagreement payoff, which may not always align with fair and equitable negotiation goals.

The MSE estimator,  $\hat{\theta}_{\mathrm{MSE}}$, is the most conservative estimate; however, it is important to note that it favors the party with the weaker disagreement payoff. This may align with a fair and equitable negotiation if the stronger party is willing to give up some profit to make the deal.

The bargaining models with the associated cost functions may have their unique purpose in some negotiation circumstances. However, by choosing Case 1 with $\hat{\theta}_{\mathrm{ABS}}$, the parties are assured they have the most equitable negotiation framework, thereby defining a reasonable royalty.

\section{Conclusion}

Navigating the complexities of determining an optimal royalty amidst uncertainty in economic opportunity costs within the NBS is challenging. However, by establishing upper and lower bounds for economic opportunity costs and employing estimation theory, parties are better equipped to handle the inherent risks and uncertainties of royalty negotiations. Estimation theory not only refines the negotiation process but also enhances the fairness and efficacy of the deal. Parties can utilize the Absolute-Value, Uniform, and Square Error cost functions to accommodate risk in a hypothetical negotiation. Following this approach, the parties may arrive at a more reasonable royalty, aligning with \emph{Georgia Pacific} factor fifteen.

\bibliographystyle{names}
\bibliography{ms}

\section*{Acknowledgments}
The authors sincerely thank Kelly Kryskowski for her careful review and valuable suggestions.

\end{document}